\documentclass[12pt,twoside,a4paper]{article}
\author{Jan-Markus Schwindt$^1$, Christof Wetterich$^2$}
\date{}
\title{The cosmological constant problem in codimension-two brane models}

\begin{document} 
\maketitle 

\centerline{\small\it $^1$Institut f\"ur Physik, Universit\"at Mainz, Staudingerweg 7,
55128 Mainz}
\centerline{\small\it \quad E-mail: Schwindt@thep.physik.uni-mainz.de}
\vspace{0.3cm}
\centerline{\small\it $^2$Institut f\"ur Theoretische Physik, Universit\"at
Heidelberg,
Philosophenweg 16, 69120 Heidelberg}
\centerline{\small\it \quad E-mail: C.Wetterich@thphys.uni-heidelberg.de}
\vspace{0.7cm} 

\begin{abstract}
We discuss the possibility of a dynamical solution to the cosmological constant problem
in the context of six-dimensional Einstein-Maxwell theory. A definite answer
requires an understanding of the full bulk cosmology in the early universe, in which
the bulk has time-dependent size and shape. We comment on the special properties of
codimension two as compared to higher codimensions.
\end{abstract}

The presence of extra dimensions provides a framework in which the cosmological 
constant problem can be viewed from a different perspective. The question is no longer
why the full spacetime curvature is so small, but rather why the four-dimensional (4D) 
curvature 
contains only such a small part of the total higher dimensional curvature. It
has been known for about twenty years \cite{rusha,cw1} that in a certain 
subclass of six-dimensional 
solutions, namely those with time-independent size and shape of internal space, the 4D 
cosmological constant $\Lambda _4$ is a free integration constant of the general 
solution. The
question remains why a solution with small $\Lambda _4$ should be dynamically 
selected.
Recently this problem was reconsidered in the context of codimension-two braneworlds
with conical sigularities \cite{codim2,cline2,football1,c2brane},
with a bulk stabilized by magnetic flux \cite{rdss1}.

We may view the solutions with static internal space as candidates for asymptotic
solutions for large time $t$. For the approach to these asymptotic solutions, however,
internal space is not expected to be static. The evolution of the universe in this
early period with time varying geometry will decide to which value of $\Lambda _4$
the late universe will converge. Two scenarios are conceivable: The dynamical approach
to the final value of the four-dimensional curvature (i.e. of $\Lambda _4$) may occur
very early in cosmology, with a fixed $\Lambda _4$ since. This resembles the 
inflationary universe, where the final zero (or very tiny) value of the three-dimensional
curvature is selected at a very early stage. (For the Friedmann solutions the initial
density is a free integration constant, and its value very close to the critical density
is selected during the early inflationary epoch.) 

For the second alternative, the adjustment to the final value of $\Lambda _4$ is still
going on in present cosmology. This will lead to a dynamical dark energy or quintessence
\cite{cwq}. In such a scenario the asymptotic value of $\Lambda _4$ is 
typically zero, and in present cosmology dark energy is expected to contribute of the same 
order as matter. However, potential problems arise now with time varying coupling
constants.

Particular exact dynamic solutions of the 6D Einstein-Maxwell theory have been recently
found for both scenarios \cite{fbcos}. They are, however, special in the sense 
that only the size, not the geometric shape of internal space changes with time, and 
that there is no warping. A dynamical solution of the cosmological constant problem
in early cosmology is not expected in such a restricted setting and actually not
found. An investigation of this question requires a time varying geometry and warping
within the most general class of solutions consistent with the symmetries. Recent 
progress towards an understanding of this complicated dynamics
was made by Vinet and Cline \cite{cline2}. They considered
different types of singularities, allowing for a general equation of state on the
brane. The authors find no self-tuning mechanism. However,
the results of ref. \cite{cline2} are by far not sufficient for a definite answer.
The limitation of their work is the restriction to 
small perturbations around an essentially static bulk: the ``football shaped" solution
\cite{football1}.
If the self-tuning of the 4D cosmological constant takes place in the very early
universe, there is no reason to assume that the bulk was static, or even almost static,
at that time. A general analysis should take place in a bulk with the most general
(non-perturbatively) time-dependent size and shape which is consistent with the symmetries.
The curvature may then be dynamically shifted between the 4D part, the 2D
part and the warping, implying an effective time-dependent 4D cosmological constant.

This can happen even if
there are no time-dependent sources at all, only the 6D cosmological constant and a
constant magnetic flux, as we demonstrated in ref.\cite{fbcos}: The quintessence field 
$\phi$ of the effective 4D theory is given by the radius of the internal space and
is therefore related to its curvature. On the other hand, the potential $V(\phi)$ and
the time derivatives of $\phi$ induce some curvature in the 4D world. The dynamics
of $\phi$ is nothing else but an effective description of the interaction between the
different parts (2D and 4D) of the total six-dimensional curvature. \\

The allowance for a more general geometry 
complicates our subject immensely. The ``football" shaped cosmology
was characterized by a few constants (the brane tension and the monopole number)
and one single function $\phi(t)$. Within the more general geometry, there is an infinite 
number of degrees of freedom, corresponding in the effective four-dimensional world
to an infinite set of scalar fields in the singlet representation of the symmetry group.
Below we derive the most general ansatz for the six-dimensional metric with the
symmetries of three-dimensional rotations and translations and internal U(1) isometry,
and we have also computed the corresponding field equations. 

For an investigation of cosmological solutions we aim to answer the
question: Is there a large class of initial conditions for which the bulk
comes to rest asymptotically leaving a very small or zero 4D curvature? 
To study this question, we adopt a bulk-based point of view \cite{holo}, in which
the singularities (branes) are seen as properties of the bulk geometry (such as 
the mass of a Schwarzschild black hole may be seen as an integration constant of
the vacuum geometry). This simplifies our task since only the field equations
in the bulk need to be solved.
The singularities will certainly play an important role in
the development of the bulk. It is of particular interest how bulk fields respond
to the singularities, and if a part of the energy of the fields may ``fall" into 
them (like matter falls into a black hole). Due to the high complexity of the field 
equations we have not yet achieved to
answer these questions in the present paper. The aim of the present note 
is therefore more modest, i.e. to 
clarify the problem for general bulk geometries and develop a strategy
for further investigations. In the text
we mainly refer to the Kaluza-Klein context which implies that the two internal dimensions
are compactified on a scale not much different from the Planck scale. 
Nevertheless, most of our results remain valid in braneworld
scenarios with large extra dimensions.\\

As our first general observation we note that codimension two is a very special case, 
for several reasons on which we comment below. 
We already mentioned the speciality
of codimension one in an earlier paper \cite{holo}, in particular the fact that the position
of a codimension-one brane cannot be detected by a ``test particle" in the bulk.
In contrast, for codimension two or higher the type and strength of the singularity
can be inferred from the properties of the bulk geometry. Still 
codimension two is special since there exists a type of brane that is not possible for any
higher codimension: the deficit angle brane. Using coordinates $x^\mu$ for the four
large dimensions and $\rho$ and $\theta$ (with $0 \leq\theta < 2 \pi$) as cylindric
coordinates for the internal space, such a brane, or conical singularity, can be
described in the following way:
The metric components $g_{\mu\nu}$ have well defined, finite
values at the position of such a singularity (at $\rho =0$, say), while 
$g_{\theta\theta}$ is proportional to $\rho^2$ in the vicinity of the brane. This is the
usual behavior of cylindric coordinates with radial coordinate $\rho$. The only effect
of the singularity is a deficit angle $\Delta$, which is expressed in the metric by 
the fact that 
\begin{equation}
 g_{\theta\theta}=\left ( 1-\frac{\Delta}{2 \pi}\right )\rho ^2 + O(\rho^3).
\end{equation} 
The infinite curvature at $\rho =0$ is not ``visible" from outside, by which we mean
that the curvature $R$ (and in fact any invariant formed from the Riemann tensor) 
remains finite in the limit $\rho \rightarrow 0$.
The curvature and the corresponding brane tension are of the delta function
type. The finiteness of the curvature implies that there are no attractive 
forces towards the brane, at least none with a divergent behavior.

Such a type of singularity exists only in the codimension two case. Otherwise all
singularities are not of the delta function type, i.e. they are locally ``visible" from
outside by Riemann tensor components 
that diverge as $\rho\rightarrow 0$, and hence diverging forces as one
approaches them. The reason for the existence of deficit angle branes in codimension
two is that one may ``cut out" a part (i.e. the deficit angle) of the circle described
by the $\theta$ coordinate at constant $\rho$ without inducing any curvature on it
(since it is a one-dimensional object). For $D>2$, the sphere $S^{D-1}$ described by
coordinates $\theta _\alpha$ at constant $\rho$ does have curvature, and 
``cutting out" some part of it does not work. Or, equivalently, multiplying the 
$g_{\theta\theta}$'s by a constant factor induces a change in the curvature which
diverges as one approaches $\rho=0$.

In fact, the deficit angle brane is a special type of a Kasner singularity. Consider,
for simplicity, a static vacuum singularity at $\rho=0$, i.e. all metric components
are functions of $\rho$ only. We may then normalize $g_{\rho\rho}$ to 1, and the 
most general metric (in 4+D dimensions) consistent with our symmetries (in particular
internal SO(D) isometry) is
\begin{equation}\label{kasner}
 ds^2 = -c^2(\rho)dt^2 + a^2(\rho)(dx^i)^2 + b^2(\rho)\tilde{g}_{\alpha\beta}
 (\theta)d \theta ^\alpha d \theta ^\beta +d \rho^2.
\end{equation} 
In the vicinity of the singularity, the vacuum Einstein equations admit solutions
of the form
\begin{equation}
 c \sim \rho^{p_1}, \quad a \sim \rho^{p_2}, \quad b \sim \rho^{p_3}
\end{equation}
with
\begin{equation}
 p_1 + 3 p_2 + (D-1)p_3 = p_1^2 +3 p_2^2 + (D-1)p_3^2 =1.
\end{equation}
The deficit angle brane corresponds to the very special solution with $p_1=p_2=0$ and
$p_3=1$ which exists only for $D=2$. In all other cases the $g_{\mu\nu}$ components
become irregular at $\rho=0$ (either zero or infinite), and some components of
the Riemann tensor diverge.

In the presence of bulk matter, the singularities may have an important influence on
the cosmological evolution. Except for the deficit angle case, they may attract the
matter and force it to fall into them, making the singularities grow (as it is
familiar for black holes).
As an example, consider $p_2=0$ (i.e. constant $a$) where
\begin{equation}
 p_1=\frac{1 \pm \sqrt{1+D(D-2)}}{D}, \quad 
 p_3=\frac{1}{D}\mp \frac{\sqrt{1+D(D-2)}}{D(D-1)}.
\end{equation}
For $D=3$ one has a solution with a ``black hole singularity" ($p_1=-\frac{1}{3}$, 
$p_3=\frac{2}{3}$) in internal space. One would expect the existence of solutions
where matter falls into this singularity, thereby changing the strength of the 
singularity or the associated brane tension (given by the ``mass" of the black hole).
Of course, the analogy of such a cosmological solution with a black hole is only formal.
In the effective four-dimensional theory there is no local object since the solution
is actually a direct product of time-warped internal space and flat three-dimensional
geometry. The time singularity appears only for a particular point in internal space.
Integrating over internal space may lead to a perfectly regular time in the effective
four-dimensional world.

We may take this discussion as a warning that results for singularities with 
codimension two should not be too easily generalized to higher codimension. It is well
conceivable that the strength of singularities does not change with time
for codimension two branes whereas it generically does for higher codimension.\\

A second speciality of codimension two arises when one considers the most general
metric consistent with certain symmetries:   
We want to look for cosmological solutions 
with a general shape of the two-dimensional internal space. 
(Static and de Sitter-like solutions were described in refs. \cite{rdss1,rusha,cw1}.) 
The first step of a dynamical investigation is the selection of an appropriate ansatz
for the metric. We will see that the determination of the most general metric 
consistent with the symmetries is nontrivial and actually extends beyond the metrics 
considered so far \cite{cline2}.
We require the following symmetries:
three-dimensional translation and rotation invariance, acting on the coordinates
$x^i$, and a U(1) symmetry, acting on the coordinate $\theta\in [0,2 \pi]$.   
No metric function should depend on $x^i$ or $\theta$, and no direction in the 
three-dimensional space should be preferred. (For simplicity, we will take this space 
to be flat, so that the metric components $g_{ij}$ are $a^2 (t,\rho)\delta _{ij}$.)
We have to find the most general metric consistent with these symmetries.

Isotropy forbids
metric components $g_{ti}$, $g_{\rho i}$ and $g_{\theta i}$, since these would select 
preferred directions in three-space, e.g. by the three-vector $(g_{t1},g_{t2},g_{t3})$.
The other off-diagonal metric components $g_{t \rho}$, $g_{t \theta}$ and $g_{\rho\theta}$
are allowed, as long as they are functions of $t$ and $\rho$ only.
Up to now we have identified the most general metric consistent with the symmetries as
\begin{eqnarray}\label{genmeco}
 ds^2 = -c^2(t,\rho)dt^2 &+& a^2(t,\rho)dx^i dx^i +b^2(t,\rho)d \theta ^2 
 +n^2(t,\rho)d \rho ^2 \\ &+& 2w(t,\rho)dt d \rho +2u(t,\rho)dt d \theta +2v(t,\rho)
 d \rho d \theta . \nonumber
\end{eqnarray}  
The next step is to look how far this line element can be simplified by a coordinate
transformation. Therefore one has to find the possible transformations consistent
with the symmetries, which should still be represented by the new coordinates
${x^i}'$ and $\theta '$. 
Transformations can 
never depend on $\theta$, since this would lead to metric 
functions depending on $\theta '$;
for example $t \rightarrow t '=t+ \delta t(\theta)$, $\theta \rightarrow \theta ' 
=\theta$ would imply $t=t'-\delta t (\theta ')$, and so $f(t)\rightarrow f'(t',\theta ')$
for any function $f$. Similarly, $t'$, $\theta '$ and $\rho '$ cannot depend on $x^i$.
Furthermore, for $\theta\rightarrow\theta '$, one has
\begin{equation}
{g^{\theta '\theta '}}=\left ( \frac{\partial\theta '}{\partial\theta}\right ) ^2 
g^{\theta\theta}, 
\end{equation}
and we impose $\partial\theta '/\partial\theta =1$, since $\theta '$ should be 
in the interval $[0,2 \pi]$. 

Transformations of $x^i$ cannot depend on $t$ or $\rho$,
since this would lead to forbidden components via
\begin{equation}
 {g^{t'i'}}=\frac{\partial t'}{\partial t}\frac{\partial {x^i}'}{\partial t}g^{tt},
\end{equation}
and similarly for $g^{\rho i}$.
Obviously, the only effect of a transformation $x^i \rightarrow {x^i}'(x^j)$ could be 
a rescaling of three-dimensional space, so we can forget about them in this context.
We are left with the following possibilities:
\begin{eqnarray} \label{symtra}
 x^i & \rightarrow & x^i ,\quad\quad
 \theta \rightarrow  \theta + \delta (t,\rho),\\ \nonumber
 t & \rightarrow & t'(t,\rho) ,\quad\quad
 \rho \rightarrow  \rho '(t,\rho).
\end{eqnarray}
There are three
off-diagonal metric components, $g_{t \rho}$, $g_{t \theta}$ and $g_{\rho\theta}$,
and one might think that these can be removed by the three remaining coordinate
transformations. It turns out that this is in general not true. The reason for that is 
essentially the U(1) symmetry. (In fact, the metric {\it can} always be diagonalized, but
then in general the new coordinate $\theta '$ will not reflect the U(1) symmetry any 
more, and fields will depend on $\theta '$.) To see this, consider the inverse of the 
metric. The components $g_{t \theta}$ and $g_{\rho\theta}$ will be zero if and only if
$g^{t \theta}$ and $g^{\rho\theta}$ are zero. The condition that this happens after
a coordinate transformation of the type (\ref{symtra}) is
\begin{eqnarray}
 {g^{t' \theta '}} &=& \frac{\partial t'}{\partial t} \left ( g^{t \theta}
  + \frac{\partial\theta '}{\partial t}g^{tt}+\frac{\partial\theta '}{\partial\rho}
  g^{t \rho}\right ) 
   + \frac{\partial t'}{\partial\rho}\left ( g^{\rho\theta}
  + \frac{\partial\theta '}{\partial t}g^{\rho t}+\frac{\partial\theta '}{\partial\rho}
  g^{\rho\rho}\right ) =0, \\
  {g^{\rho '\theta '}} &=& \frac{\partial\rho '}{\partial t} \left ( g^{t \theta}
  + \frac{\partial\theta '}{\partial t}g^{tt}+\frac{\partial\theta '}{\partial\rho}
  g^{t \rho}\right ) 
   + \frac{\partial\rho '}{\partial\rho}\left ( g^{\rho\theta}
  + \frac{\partial\theta '}{\partial t}g^{\rho t}+\frac{\partial\theta '}{\partial\rho}
  g^{\rho\rho}\right ) =0.
\end{eqnarray}
A solution of these differential equations implies either that the Jacobi determinant
of the $(\rho ,t)$ transformation vanishes,
\begin{equation}
 det \left ( \begin{array}{cc} \frac{\partial t'}{\partial t} & 
 \frac{\partial t'}{\partial\rho} \\ \frac{\partial\rho '}{\partial t}
 & \frac{\partial\rho '}{\partial\rho} \end{array}\right ) =0,
\end{equation}
which is not possible, or that the brackets vanish. But the second possibility consists
of two conditions for the function $\theta '$, which can in general not be fulfilled 
simultaneously. 

One concludes that generally only one of the two components $g^{t \theta}$ and
$g^{\rho\theta}$ can be set to zero (in contrast to \cite{cline2}). 
A procedure to simplify the metric (\ref{genmeco})
could look as follows: Use the freedom for $t'$ and $\rho '$ to annihilate
$g^{t \rho}$ and for one further simplification, e.g. to arrange that 
${g_{tt}}'=-{g_{ii}}'$, i.e. to make time conformal with respect to space. 
Then use the freedom for $\theta '$ to annihilate either
$g^{t \theta}$ or $g^{\rho\theta}$. The simplified line element is then
\begin{equation}\label{ugauge}
 ds^2 = a^2(t,\rho)(-dt^2 + dx^i dx^i) +b^2(t,\rho)d \theta ^2 + n^2(t,\rho)d \rho ^2
 + 2 u(t,\rho)dt d \theta ,
\end{equation}
or similarly with $2 v(t,\rho)d \rho d \theta $ instead of $2 u(t,\rho)dt d \theta$.
In the effective four-dimensional picture $u$ corresponds to the time component of an
abelian gauge field (hence some kind of electric potential), since $g_{\theta\mu}$
integrated over internal space is the gauge field corresponding to the U(1) isometry.
On the other hand $v$ corresponds to a scalar field. The fact that a degree of freedom 
can be shifted between a scalar field and the component of a gauge field is a familiar
fact in particle physics.

The presence of off-diagonal metric components $g_{t \theta}$, $g_{\rho\theta}$, which
cannot be transformed away simultaneously, is a special feature of codimension-two
models. Consider $D>2$ internal dimensions, 
and $D-1$ of these dimensions, represented by 
coordinates $\theta _\alpha$, were symmetric under, say, SO(D), then the
$g_{t \theta}$ and $g_{\rho\theta}$ components would be forbidden, because they would
select preferred directions in the $D-1$-dimensional space,
in conflict with the SO(D) symmetry. The difference is that
a U(1) ``rotation" is a translation rather than a rotation. In this sense a 
codimension-two spacetime is more complicated than a higher-dimensional one.\\

A third special feature of codimension two is that internal space can be compactified
and stabilized by a gauge field $A_B$ in a monopole configuration (capital
indices run over all six dimensions). Six-dimensional Einstein-Maxwell theory \cite{rdss1}
according to the action
\begin{equation}
 S=\int d^6 x \sqrt{-g}\left \{ -\frac{M_6^4}{2}R + \lambda _6 + \frac{1}{4}F^{AB}F_{AB}
 \right \},
\end{equation}
is a convenient toy model for higher dimensional scenarios. Here $M_6$ is the
reduced Planck mass corresponding to six-dimensional gravity, $\lambda _6$ is a
cosmological constant term and $F_{AB}$ is the field tensor of the gauge field.
Including the gauge field into our considerations, we find that   
the three components $A_t$,
$A_\rho$ and $A_\theta$ are allowed by the symmetries. One can choose to set either
$A_t$ or $A_\rho$ to zero by a gauge transformation. This is similar to the choice
between $g^{t \theta}$ and $g^{\rho\theta}$ described above.

Comparing this cosmological Einstein-Maxwell system to the static case, one finds that the
ordinary differential equations (containing only $\rho$-derivatives) 
are generalized to partial differential equations,
containing $t$- and $\rho$-derivatives. The three functions $a$, $b$ and
$A_\theta$, which are already present in the static case, 
are accompanied by three more functions: $n$, $u$ or $v$, and $A_t$ or $A_\rho$. 

We have computed the field equations for the six independent functions of $t$ and $\rho$.
As an example we give the $(tt)-$ component of Einstein's equations, in the gauge
$a=c$ and $w=v=0$:
\begin{eqnarray}
G^t_t &\equiv& -\frac{1}{a^2(1+q^2)}\left ( 3 \frac{\dot{a}^2}{a^2}+ 3 \frac{\dot{a}\dot{b}}
 {ab} + 3 \frac{\dot{a}\dot{n}}{an} +\frac{\dot{b}\dot{n}}{bn} \right ) 
 + \frac{1}{n^2} \left ( 3 \frac{{a'}^2}{a^2} -3 \frac{a'n'}{an}
 +3 \frac{a''}{a} \right ) \\
 \nonumber &+& \frac{1}{n^2 (1+q^2)}\left ( 3 \frac{a'b'}{ab} -\frac{b'n'}{bn}
 - \frac{n'u'q^2}{2nu} +\frac{b''}{b}+\frac{u''q^2}{2u} \right )\\
 \nonumber &+& \frac{q^2}{n^2(1+q^2)^2}\left ( \frac{a'b'}{ab}+\frac{{b'}^2}{b^2}
 +\frac{a'u'}{au}(1+\frac{3}{2}q^2)-\frac{3b'u'q^2}{2bu}+\frac{{u'}^2}{4u^2}(1-q^2)\right )\\
\nonumber &=& 8 \pi G_6 
T^t_t \equiv \frac{4 \pi G_6}{1+q^2}\left ( -\frac{\dot{A}_\theta ^2}{a^2 b^2}
 -\frac{{A'_t}^2}{a^2 n^2}-\frac{{A'_\theta}^2}{n^2 b^2} \right ).  
\end{eqnarray}
Here dots and primes denote derivatives with respect to $t$ and $\rho$, respectively,
and we use the abbreviation $q^2 \equiv u^2 /(a^2 b^2)$. 
The equations for the other components are of similar length and are not displayed
here.
A full numerical analysis of this system would involve as initial conditions twelve
functions of $\rho$ (four metric and two gauge field components and their first time
derivatives at some initial time $t_0$) which are subject to three constraint
equations, namely the $(tt)-$, $(t \rho)-$ and $(t \theta)-$ components of Einstein's
equations, which contain no second time derivatives. 
The time evolution is determined by the $(ii)-$, $(\theta\theta)-$,
$(\rho\rho)-$ and $(\theta\rho)$- components of Einstein's equations and two equations
for the gauge field.

Again we want to compare this to an Einstein-Maxwell system in higher codimensions.
We already showed that the metric components $g_{t \theta _\alpha}$ and
$g_{\rho\theta _\alpha}$ are not consistent with a symmetry larger than U(1) acting
on the $\theta$ coordinates.
For the gauge field the situation is slightly different. For specific solutions (solitons)
a component $A_{\theta _\alpha}$ may be allowed 
even if the internal symmetry is larger than 
U(1). An example is the monopole solution on $S^2$. Although $A_\rho =0$ and 
$A_\theta \neq 0$, the $\theta$-direction is not
preferred physically. A coordinate transformation may be accompanied by a gauge
transformation, so that the transformed $A$-field lies in the new $\theta$-direction.
An analogous procedure does not work for the metric tensor, since the gauge
transformations are the coordinate transformations themselves.
But the components $A_\rho$ and $A_t$ are not necessary in the $D>2$ case: Without the
$g_{t \theta}$ and $g_{\rho\theta}$ metric components, the components ${G^t}_\theta$
and ${G^\rho}_\theta$ of the Einstein tensor are identically zero. The corresponding
components of the energy momentum tensor induced by the Maxwell field,
\begin{equation}
 T_{AB}^{(F)}=F_{AC}{F_B}^{C}-\frac{1}{4}F_{CD}F^{CD}g_{AB},
\end{equation}
are then also zero,
which implies $F_{\rho t}=0$ (as long as $A_{\theta _\alpha}$ is non-trivial) 
and so $A_\rho$ and $A_t$ are pure gauge. 
Compared to the codimension-two case, the higher codimensions therefore involve 
two functions less:
Only $a$, $b$, $n$ and $A_\theta$ remain after appropriate simplifications.\\

We already mentioned that, after removing two of the off-diagonal metric components,
there remains still one degree of freedom for coordinate changes in order to bring the
metric into a pleasant form. We chose to use this freedom to make time conformal, i.e.
$-g_{tt}=g_{ii}$. We warn, however, that this may often not be the convenient choice, 
because the corresponding time coordinate
may be different from the ``physical'' time. To explain this, remember that in usual
four-dimensional cosmology, time can be made conformal by a transformation $t \rightarrow
\tau (t)$. In the six-dimensional model, we need instead transformations $t,\rho
\rightarrow t'(t,\rho),\rho' (t,\rho)$ in order to bring the metric into the 
required form. 
This may mix the time and $\rho$ coordinate to some extent.
The effective four-dimensional 
Lagrangian is obtained by integrating out internal space in the form
\begin{equation}
 L_{eff}(t,x^i)\sim\int d \rho d \theta \sqrt{g_{int}} L(t,x^i,\rho,\theta),
\end{equation}  
or similar, where $g_{int}$ is the determinant of the metric of the internal space.
This effective Lagrangian obviously depends on the choice of the $(t,\rho)$ 
frame one uses. Nevertheless, all the effective Lagrangians derived from 
different choices of $t$ and $\rho$ must describe the same physics because they
are obtained from the same six-dimensional theory. Such an equivalence between 
different Lagrangians will in general not be seen easily, because there is an infinite
number of fields mixed with each other when going from one frame to the other.

To illustrate this, we consider the example of five-dimensional Kaluza-Klein
theory with action
\begin{equation}
 S=-\frac{1}{16 \pi G_5}\int d^5 x \sqrt{-g}\, R.
\end{equation} 
The four large dimensions are again parametrized by coordinates $x^\mu$,
and the fifth coordinate $y$ runs from 0 to $2 \pi r$. If one writes the metric in 
the form
\begin{equation}
 g_{AB}=\phi^{-1/3}\left ( \begin{array}{cc}
 \tilde{g}_{\mu\nu}A_\mu A_\nu \phi & A_\nu \phi \\
 A_\mu \phi & \phi
 \end{array}\right ),
\end{equation}
integration over $y$ leads to the following 4D action for the zero modes:
\begin{equation}\label{kk5d}
 S_{eff}^{(0)}=\frac{1}{16 \pi G_4}\int d^4 x \sqrt{-\tilde{g}}\left (
 -\tilde{R}^{(0)}+\frac{1}{4}\phi^{(0)}F_{\mu\nu}^{(0)}F^{\mu\nu (0)}
 -\frac{1}{6}\frac{\partial _\mu \phi^{(0)}\partial^\mu \phi^{(0)}}{(\phi^{(0)})^2}
 \right ).
\end{equation}
Here $G_4=G_5/2 \pi r$, $F_{\mu\nu}=\partial _\mu A_\nu -\partial _\nu A_\mu$, the
superscript $(0)$ denotes that in the Fourier expansion of components with respect to 
$y$ only the zero modes are taken into account, and a tilde denotes a quantity
constructed from $\tilde{g}_{\mu\nu}$.

Now we perform a small coordinate transformation, affecting only the time coordinate
in the form $t \rightarrow t'=t-\epsilon\,\sin \frac{y}{r}$. The corresponding
change of the metric is (to order $\epsilon$)
\begin{eqnarray}
 g'_{ty}(x^i,t',y)&=& g_{ty}(x^i,t(t',y),y)+\epsilon \;\frac{g_{tt}(x^i,t(t',y),y)}{r}
 \cos \frac{y}{r}, \\
 g'_{yy}(x^i,t',y)&=& g_{yy}(x^i,t(t',y),y)+ 2 \epsilon \;\frac{g_{ty}(x^i,t(t',y),y)}{r}
 \cos \frac{y}{r}.
\end{eqnarray}
This transformation does not only mix the fields $\phi$, $A_\mu$ and $\tilde{g}_{\mu\nu}$
in a complicated way. It also changes the line of $y$-integration defined by $x^\mu =const$.
In the transformation of the fields this is reflected by additional time derivatives, e.g.
\begin{equation}
 \phi '=\phi \;\left (1 + 3 \epsilon \; \frac{A_t}{r}\cos\frac{y}{r}+ \epsilon \;
 \frac{\dot{\phi}}{\phi}\sin\frac{y}{r} \right ).
\end{equation}
The 4D action for the zero modes of the new fields will again be eq.(\ref{kk5d}), 
but these zero modes are not only combinations of the old zero modes. They contain
admixtures of higher Fourier modes of the original fields (due to the $\cos
\frac{y}{r}$ term) and even of their time derivatives.

For a given solution of the field equations,
there are certainly choices of
coordinates in which the four-dimensional world looks simpler than 
in others. In some situations there may be a clear and unique preferred frame 
which identifies a ``physical" time coordinate.
The ``physical'' time is easily identified when there is a timelike Killing vector. 
Returning to the six-dimensional model, such a Killing vector is given 
for the Kasner solutions mentioned before. 
In a situation like the static football shaped solution ($p_1=p_2=0$ at the singularities),
where time and three-space are not differently warped, an appropriate frame has
necessarily $-g_{tt}= g_{ii}$. 
For Kasner solutions
with $p_1 \neq p_2$, like the aforementioned black hole type singularities, time and 
three-space are warped differently, and
the physical time (with time axis parallel to the Killing vector) 
corresponds to a frame with $-g_{tt}\neq g_{ii}$. 
In this frame all metric components depend only on $\rho$ (cf eq (\ref{kasner})).
In general, Lorentz invariance will be broken in the effective 4D world corresponding
to such a solution. By suitable $(t,\rho)$
transformations one still finds local charts with $-g_{t't'}= g_{ii}$, 
but these coordinates
do not represent the symmetry of the solution, since now functions depend on $t'$ and
$\rho'$: $f(\rho)=f(\rho(t',\rho'))$. Everything would look unnecessarily
complicated in such a frame, which is therefore ``unphysical".

Even for late cosmology, say the present epoch,
there is no exact timelike Killing vector, only an 
approximate one. This means that for some choice of coordinate frame variables
vary only very slowly with time. The identification of the ``correct" time 
is more complicated in such a situation than in those with exact
Killing vector. For a wrong choice of $(\rho ,t)$-frame variables may vary much too fast
with time and the geometry seems to get totally distorted.
This is similar to (but worse than) the unphysical gauge modes
appearing in some approaches of cosmological perturbation theory 
(e.g. in synchronous gauge).
The identification of the time coordinate relevant for the effective
four-dimensional physics is a serious task in higher-dimensional cosmology.
Conclusions based on the ``conformal gauge" ($a=c$ in eq.(\ref{genmeco}))
can easily be misleading. In practice, this means that it may be advisable to work
with a metric that is even more general than the ansatz (\ref{ugauge}).\\

In summary, we have computed the field equations for the six-dimensional Einstein-Maxwell
theory for the most general ansatz of the metric and gauge fields consistent with
the symmetries of three dimensional rotations and translations and a U(1)-isometry.
A crucial issue for a possible dynamical solution of the cosmological constant
problem is the possibility that the brane tension changes with time. For a restricted
ansatz it was found that this does not happen for this system \cite{codim2},
but we would like to emphasize that a complete answer needs the most general ansatz
for the metric. Therefore a
dynamical solution to the cosmological constant problem
in the context of six-dimensional brane or Kaluza-Klein models is so far not ruled out,
not even in the case of infinitely thin deficit angle branes.
An answer to the question requires a much more detailed understanding of the early 
universe dynamics with a time-dependent bulk geometry. 
As we have shown, such an understanding is
complicated by the fact that the four-dimensional interpretation of the six-dimensional
dynamics is far from clear in the absence of a timelike Killing vector.

Furthermore, we have pointed out that codimension
two is a very special case for several reasons:\\
(i) There exist conical singularities (deficit angle branes) which do
not induce any attractive forces in the bulk.
(ii) The metric is relatively complicated even if one requires that all quantities
depend only on one internal coordinate $\rho$ and on time.
(iii) A Maxwell field can compactify and stabilize the internal space. 

We have argued by analogy with the dynamical black hole geometries with infalling 
matter in the four-dimensional world that in general the strength of a singularity
can vary with time. Once such time varying ``brane tensions" can be desribed 
properly in a higher-dimensional world the issue of the cosmological constant or 
quintessence may show new, unexpected facets.

\end{document}